\documentclass[twocolumn]{aa}
\usepackage{graphicx}
\usepackage{txfonts}
\usepackage{natbib,multirow}
\bibpunct{(}{)}{;}{a}{}{,}
\begin{document}
   \title{The infrared L'-band view of the Galactic Center \\ with NAOS-CONICA at VLT\thanks{Based on observations collected at the European Southern Observatory, Paranal, Chile}}

   \author{Y. Cl\'enet\inst{1}
\and
D. Rouan \inst{1}
\and
E. Gendron \inst{1}
\and
F. Lacombe \inst{1}
\and
A.-M. Lagrange \inst{2}
\and
G. Rousset \inst{3}
\and
R. Genzel \inst{4}
\and
R. Sch\"odel \inst{4}
\and
T. Ott \inst{4}
\and
A. Eckart \inst{5}
\and
O. Marco \inst{6}
\and
L. Tacconi-Garman \inst{7}}

\offprints{Y. Cl\'enet, \email{yann.clenet@obspm.fr}}

   \institute{Observatoire de Paris, LESIA, 5 place Jules Janssen, F-92195 Paris Cedex, France\\
              \email{firstname.lastname@obspm.fr}
         \and
             Observatoire de Grenoble, LAOG, BP 53, F-38041 Grenoble Cedex 09, France\\
             \email{firstname.lastname@obs.ujf-grenoble.fr}
         \and
             Office National d'Etudes et de Recherches A\'erospatiales (ONERA), DOTA, BP 72, 92322 Ch\^atillon Cedex, France\\
              \email{rousset@onera.fr}
         \and
             Max-Planck-Institut f\"ur extraterrestrische Physik, Giessenbachstrasse, D-85748 Garching, Germany\\
             \email{lastname@mpe.mpg.de}
         \and
             I. Physikalisches Institut, Universit\"at zu K\"oln, Z\"ulpicher Strasse 77, D-50937 K\"oln, Germany\\
             \email{eckart@ph1.uni-koeln.de}
         \and
             European Southern Observatory (ESO), Alonso de C\'ordova, Casilla 3107, Vitacura, Casilla 19001, Santiago 19, Chile\\
             \email{omarco@eso.org}
         \and
             European Southern Observatory (ESO), Karl-Schwarzschild-Strasse 2, D-85748 Garching bei M\"unchen,
Germany\\
             \email{ltacconi@eso.org}}

   \date{Received 16 April 2003; accepted}

   \abstract{ We report on Galactic Center L'-band observations made during NAOS/CONICA Science Verification. Colors of the inner 2\arcsec\ stars reveal an infrared excess of \object{S2}, the closest star to the black hole, that could sign the first thermal infrared detection of \object{Sgr A*}. A multi-wavelength maximum likelihood analysis has allowed us to eliminate all but two of the candidates for gravitational lensing proposed by \citet{alexander01}. Our observations of the thin and intersecting structures of the Northern Arm could trace several shocks heating the neighbooring dust rather than a stream of  matter in orbit around the central mass as previously thought.

   \keywords{Galaxy: center, stellar content - Infrared: stars - Instrumentation : adaptive optics - Stars: imaging
               }
   }

   \maketitle
\titlerunning{The L'-band view of the GC with NACO}
\authorrunning{Y. Cl\'enet et al.}

\section{Introduction}
In the infrared (IR), our knowledge of the central parsecs of the Galactic Center (GC) has highly benefited from the improvements  in high spatial resolution techniques. Thanks to speckle imaging and later adaptive optics (AO) at ESO and Keck, two groups at MPE and UCLA have been able, through  proper motion measurements of the central stars, to severely constrain the nature/value of the central compact mass of our Galaxy \citep{schodel02,ghez03a}.

Thermal IR observations at high angular resolution are of prime interest for the search of an IR counterpart of \object{Sgr~A*} since this wavelength domain still benefits from the spatial resolution needed to separate the individual sources of the \object{Sgr~A*} cluster. Furthermore, the sensitivity achieved by NAOS/CONICA (hereafter NACO, \citealp{rousset00,lenzen98}) in the thermal IR should allow one to reach the flux predicted by recent \object{Sgr~A*} emission models \citep{markoff01,yuan03}, whereas these predictions drop at shorter wavelengths. Previous thermal IR studies \citep{depoy91,simons96,blum96,clenet01} had a too low sensitivity and/or spatial resolution to detect \object{Sgr A*} and were concentrated on the analysis of the bright stars (m$_\mathrm{L}$$<$11), deriving classification of the more isolated ones.

The opening of a new window in the L'-band with NACO gives the opportunity to perform multi-wavelength IR studies that have only been tackled in the H- and/or K-bands till now (eg gravitational lensing by \object{Sgr~A*},  ionized gas free free emission, dust emission or bow-shocks).

%__________________________________________________________________

\section{Observations and data reduction}
Observations of the GC have been performed with the 8 meter VLT UT4 telescope (ESO) during NACO Science Verification (SV). 1024$\times$1024 pixel images have been acquired with the IR CONICA camera, after correction from atmospheric turbulence by the NAOS AO system. 

On 2002 August 29, H- ($\lambda_\mathrm{c}$=1.66 $\mathrm{\mu m}$) and Ks-band ($\lambda_\mathrm{c}$=2.18 $\mathrm{\mu m}$) images have been obtained using the IR wavefront sensor (WFS), servoed on the bright K-band source \object{IRS~7} (m$_\mathrm{K}$=6.5) which is located about 5.5\arcsec\ north to \object{Sgr A*}. To date, NAOS is the only astronomical AO system equipped with an IRWFS, allowing correction even when no visible counterpart is available for the classical visible WFS. The pixel scale was 0.01326\arcsec/pixel. 

\begin{figure}
\centering
\begin{tabular}{c}
\resizebox{\hsize}{!}{\includegraphics{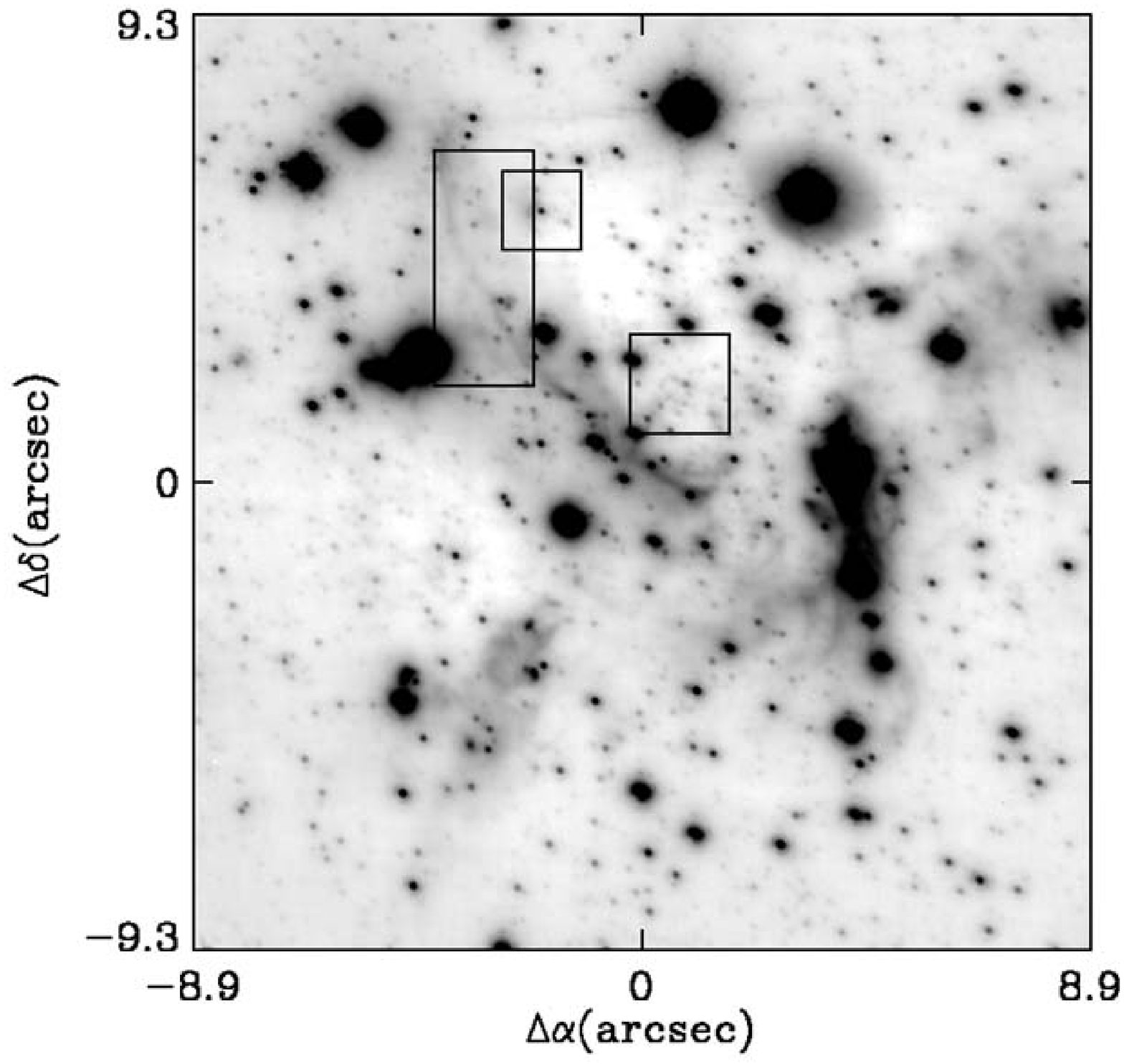}}\\
\resizebox{\hsize}{!}{\includegraphics{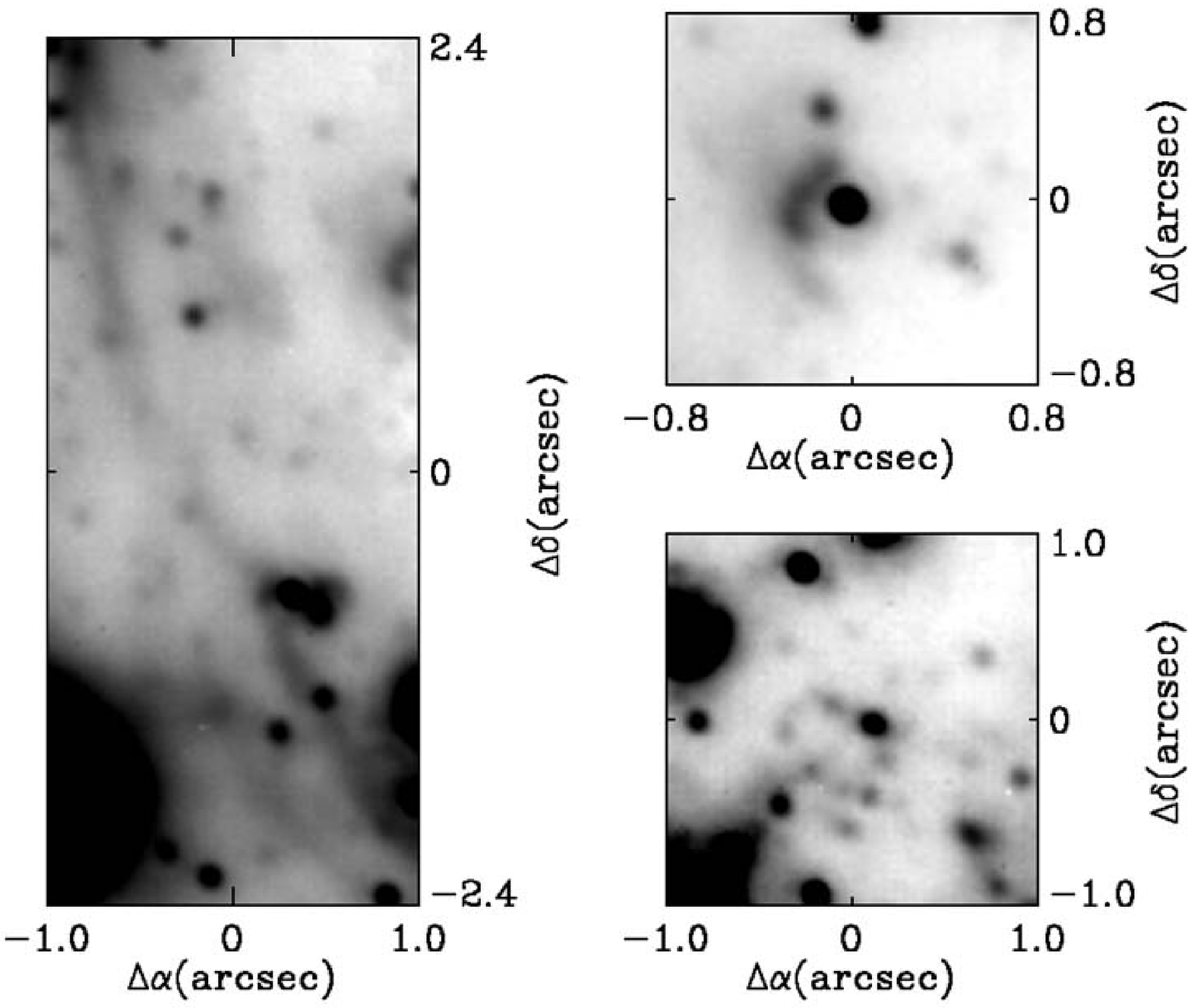}}\\
\end{tabular}
\vspace{-5mm}
\caption{Up: NACO L'-band image of the GC. Flux is displayed on a power-law scale (I$^{0.7}$) to highlight the faint extended emission structures. Locations of zooms are shown by rectangles. Down left: Zoom on the Northern Arm (I$^{0.7}$ power-law scale). Down right: Zoom on the new detected bow shock (upper image) and on the \object{Sgr~A*} cluster (lower image).}
\label{fig:imagelp}
\end{figure}

For operational reasons, the visible WFS was the only one available for the SV L'-band ($\lambda_\mathrm{c}$=3.80 $\mathrm{\mu m}$) observations (2002 August 19) and the AO reference star was the one usually chosen for visible WFS (m$_\mathrm{R}$=13.8), located about 25\arcsec\ north-east to \object{Sgr A*}. As a result, the stars appear elongated toward the direction of the reference star (Fig.~\ref{fig:imagelp}). The pixel scale was 0.0271\arcsec/pixel.

In each band, a data cube has been created from the acquisition, in a random jitter mode (8\arcsec\ jitter box width at H and Ks, 10\arcsec\ at L'), of 25 images at H, 20 at Ks and 76 at L'. Each individual image of these data cubes results from the coaddition of 4 exposures of 15~s at H and Ks, 150 exposures of 0.2~s at L'. In each band, the data reduction operations were the following:

\begin{enumerate}
\item L'-band images only: to correct from the large-scale variations of the sky background, subtraction for each data cube image of the median value computed over the entire FOV
\item for each data cube image, subtraction of the sky emission map built as a median along the third direction of the data cube
\item flat field division to correct from the instrumental response
\item  bad pixels correction
\item selection of images according to the estimated Strehl ratio, evaluated from the central flux of a star observed on each individual exposure. 8 images (out of 76) have been kept in the L'-band, 18 in the H-band and 14 in the Ks-band.
\item recentering of each individual selected image
\item coaddition of the common part of each individual image, after a background adjustment in order to put the least bright regions  to a value near zero.
\end{enumerate}

The resulting L'-band image is displayed in Fig.~\ref{fig:imagelp}, together with a zoom showing a newly observed bow shock located around 3.4\arcsec\ north and 2.8\arcsec\ west to \object{Sgr A*}, a close up of the Northern Arm and a zoom on the \object{Sgr A*} cluster. H- and Ks-band SV images are not presented here since they are discussed in \citet{genzel03a}.

\begin{table*}
\centering \caption[]{Dereddened colors of the \object{Sgr~A*} cluster stars (Fig.~\ref{fig:imagelp}). IRS~16 stars are not included. Offsets (in unit of 10 mas) are relative to \object{S2}.}
\label{table:color}
\resizebox{\hsize}{!}{\begin{tabular}{l c c c c c c c c c c c c c c c c c c c c c c}
   \hline
   \hline
   ID Number & 1 & 2 & 3 & 4 & 5 & 6 & 7 & 8 &9 & 10 & 11 & 12 & 13 & 14 & 15 & 16 & 17 & 18 & 19 & 20 & 21 & 22 \\
\hline
   Other name & & & & \object{S7} & & \object{S8} & & \object{S4} & \object{S9} & \object{S11} & & \object{S10} & \object{S2} & \object{S1} & \object{S12} & & & & & & & \\ 
\hline
  $\Delta\alpha$ & 96 & 76 & 52 & 47 & 39 & 35 & 31 & 24 & 18 & 16 & 15 & 2& 0 & -6 & -11 & -31 & -50 & -55 & -59 & -67 & -81 & -85\\
\hline
  $\Delta\delta$ & 1 & -69 & -43 & -2 & 85 & -25 & -93 & 12 & -35 & -57 & 48 & -39& 0 & -21 & 27 & 27 & -18 & -64 & 37 & -88 & -29 & -102\\
\hline
   (H-Ks)$_0$ & 0.4 & 0.4 & 0.5 & 0.5 & 0.4 & 0.4 & 0.7 & 0.4 & 0.5 & 0.3 & 0.5 & 0.6& 0.4 & 0.4 & 0.4 & 0.4 & 0.8 & 0.3 & 0.6 & 0.4 & 0.5 & 0.7\\ 
\hline
   (Ks-L')$_0$ & -0.5 & 0.0 & -0.1 & 0.3 & -0.2 & -0.1 & 0.0 & -0.1 & 0.1 & 0.1 & 0.3 & -0.2 & 0.8 & -0.3 & 0.5 & 0.2 & 2.3 & 3.7 & -0.1 & 0.2 & -0.5 & -0.3\\
\hline
   L' & 12.1 & 11.0 & 11.9 & 13.3 & 10.8 & 13.0 & 11.1 & 12.8 & 13.4 & 12.7 & 13.7 & 12.8 & 11.5 & 13.4 & 13.4 & 14.0 & 13.6 & 11.9 & 12.9 & 12.4 & 12.6 & 12.4 \\
\hline
\end{tabular}}
\end{table*}

In the three bands, the relative photometry has been computed with Starfinder \citep{diolaiti00}.  To calibrate the photometry, we have firstly selected from \citet{ott99} a number of stars known for their non variability. In the L'-band, we have extracted from the reference stars list presented in Fig.~3 of the aforementioned article the objects whose photometry was available in \citet{blum96}: \object{IRS~29N}, \object{IRS~33SE}, \object{IRS~16C}, \object{IRS~16CC}, \object{IRS~21} and \object{MPE+1.6-6.8}. In the Ks-band, almost all these reference stars were saturated. We have used the photometry of all non saturated stars given in Table~2 of the same article which matched the non variability selection criterion given by the authors: $\chi^2$$<$5.13 \citep[see][ for a definition of $\chi^2$]{ott99}: \object{ID 124}, \object{137}, \object{144}, \object{148}, \object{151}, \object{156}, \object{162}, \object{163}, \object{179}, \object{195}, \object{198}, \object{212}. In the H-band, some of the reference stars of \citet{ott99} were also saturated and we have extracted from the same Table~2 the non saturated stars whose photometry was available in \citet{blum96}: \object{IRS~33SE}, \object{IRS~16CC}, \object{IRS~33SW}, \object{IRS~33N}, \object{MPE+1.0-7.6}.

The mean zero point of each band has been then computed by considering the list of non variable stars retained as explained above. The corresponding standard deviations give an estimation of the photometric errors: 0.28 in the H-band, 0.07 in the Ks-band and 0.35 in the L'-band, since photometric errors computed by Starfinder are much lower than these values.

One should note that the photometry from \citet{ott99} is given in the K-band whereas the NACO filter was Ks. To establish a new system of faint near-IR standard stars, \citet{persson98} have surveyed more than 60 standard stars and nearly 30 red stars. For the former, they have found $|$m$_\mathrm{K}$-m$_\mathrm{Ks}|$$\leq$0.025 and $|$m$_\mathrm{K}$-m$_\mathrm{Ks}|$$\leq$0.093 for the latter (reached for the very red object \object{Oph~N9}). These values being below our photometric errors, we make no difference in the following between K and Ks magnitudes. Note that the low standard deviation value found to establish Ks-band magnitudes (0.07) demonstrates that the scatter between K and Ks magnitudes is neither significant here. 

On UKIRT web pages ({\em www.jach.hawaii.edu/JACpublic/ UKIRT/astronomy/standards.html}), L- and L'-band photometries are available for about 130 standard bright stars. The magnitude difference is maximum ($|$m$_\mathrm{L}$-m$_\mathrm{L'}|$=1.3) for late-type K or M stars, with a mean value for these two classes (32 stars) of $\langle$m$_\mathrm{L}$-m$_\mathrm{L'}\rangle_\mathrm{KM}$=0.04 and a standard deviation of 0.03. Among these 32 stars, the supergiants have the same mean value and standard deviation. Then, after a comparison with the standard deviation value of the zero point estimation in the L'-band, we assume no difference between the photometry in the L-band, from \citet{blum96}, and in the L'-band (used with NACO).

\section{The IR counterpart to \object{Sgr~A*}}
On the 2\arcsec$\times$2\arcsec\ L'-band image of the \object{Sgr~A*} cluster, more than 20 stars are detected. Their dereddened colors, assuming the extinction law from \citet{rieke85}, a distance to the GC of 8 kpc \citep{reid93} and an extinction value $A_\mathrm{K}$=2.7 \citep{clenet01}, are given in Table~\ref{table:color}. Together with \object{ID\#17} and \object{ID\#18} (a possible unresolved double source as may indicate its large elongated shape), \object{S2}, the closest star to \object{Sgr A*}, has very red colors. Is this \object{S2} IR excess intrinsic to the star or due to a contribution from \object{Sgr A*} ?

The latter hypothesis cannot be directly ruled out since the distance between \object{S2} and \object{Sgr~A*} in 2002 (about 0.04\arcsec) was lower than the L'-band NACO spatial resolution (0.096\arcsec), making impossible to spatially separate the \object{S2} emission from the putative one of \object{Sgr~A*}. If we assume that \object{S2} has a (Ks-L')$_0$ index among the highest of its neighbours (say (Ks-L')$_0$=0.4) and attribute the entire color excess to the L'-band flux (since \object{S2} and the other surrounding stars have similar (H-Ks)$_0$ indices), we derive for the extra source m$_\mathrm{L'}$=12.8 and a dereddened flux density of F$_\mathrm{\nu}$=7~mJy. This is in good agreement with predictions made by both jet (rather in a flare state) and accretion disk emission models of \object{Sgr~A*} \citep{markoff01,yuan03}. Then, if confirmed, this detection could not help discriminating between these two models still in competition to explain the black hole emission, but would be the first one in the thermal IR.

\begin{figure}[h]
\resizebox{\hsize}{!}{\includegraphics{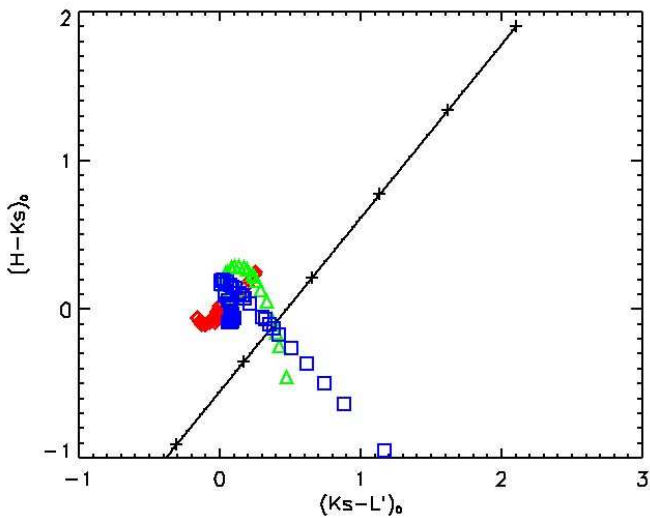}}
\caption{Dereddened color-color diagram for \object{S2}, superimposed on intrinsic colors of different classes of stars. The line represents the variation of the dereddened \object{S2} colors with respect to the interstellar extinction, from A$_\mathrm{K}$=0 (top) to A$_\mathrm{K}$$\approx$5 (bottom), and crosses on this line are for integer values of A$_\mathrm{K}$. Additional symbols are for intrinsic colors of stars from \citet{ducati01}: red diamonds are for main sequence stars, green triangles for giants and blue squares for supergiants.}
\label{fig:colS2}
\end{figure}

Is the high (Ks-L')$_0$ color index of \object{S2} compatible with intrinsic colors of a star ? \object{S2} colors are not compatible with any main sequence star indices but could match the intrinsic colors of a giant (M4 or M4.5) or a supergiant (M1 or M1.5) star (Fig.~\ref{fig:colS2}). Though, it would require an appropriate extinction coefficient (A$_\mathrm{K}$=3.6), much larger than the average extinction determined by \citet{clenet01}. The visible absolute magnitude of a M1-1.5 supergiant (M$_\mathrm{V}$=-5.6, \citealp{cox00}), with the extinction law from \citet{rieke85}, a distance to the GC of 8.0~kpc and A$_\mathrm{K}$=3.6, leads to \object{S2} predicted magnitudes of (m$_\mathrm{H}$, m$_\mathrm{Ks}$, m$_\mathrm{L'}$)=(9.6, 7.8, 5.6), which are incompatible with its apparent magnitudes (15.5, 13.6, 11.5). On the other hand, the M4-4.5 giant case is compatible with these magnitudes as the corresponding visible absolute magnitude (M$_\mathrm{V}$=-0.3 for the M5 class, the closest found in \citealp{cox00}) leads to the following predicted magnitudes: (m$_\mathrm{H}$, m$_\mathrm{Ks}$, m$_\mathrm{L'}$)=(14.9, 13.1, 10.9). Though, this giant hypothesis is invalidated by recent AO spectroscopic observations made at Keck by \citet{ghez03a}: the \object{S2} K-band spectrum exhibits no CO absorption but emission lines in agreement with a hot massive main sequence star. To date, no explanation in terms of "normal" colors of a star seems to elucidate the L'-band excess of \object{S2}. However, the thermal excess of \object{S2} may be an indication of dust in the neighbourhood of this star, either as a dust disk around the star itself, either as clumps located in the vicinity of \object{Sgr~A*}. A quantitative model is given in \citet{genzel03a} to explain the \object{S2} IR excess by dust emission. Stellar variability, a last possible interpretation, is unlikely: a variation of typically 0.4 mag at K (with respect to previous published values) would have occured within a few days, which is not consistent with usual variability of a massive main-sequence star, and \object{S2} magnitudes have remained constant over the past years.

\section{Gravitational lensing}

\citet{alexander01} has shown the interest in looking for the possible gravitational lensing of background sources by the black hole. The discovery of one or several pairs of gravitational images would allow one to precisely locate \object{Sgr A*}  in the IR, an alternate solution to SiO masers detection \citep{menten97}.

Multicolor photometry adds strong constraints on this study since  the luminosity amplification factor of a gravitational image is not wavelength dependent. Using a maximum likelihood method and the NACO photometry in the H-, Ks- and L'-bands, we have estimated the amplification factor for the ten most likely image pair candidates that \citet{alexander01} found after Monte-Carlo simulations and minimal likelihood analysis. We have considered the problem where one wants to calculate the ratio R of two flux, having three different measurements of this ratio. Through maximum likelihood estimation, the solution is given by the maximum of the following sum, taken as a function of the ratio R: $-\displaystyle \sum_{i=H, Ks,L'} \left(f_{1_\mathrm{i}}-R f_{2_\mathrm{i}}\right)^2/\left(\sigma_{1_\mathrm{i}}^2+R^2 \sigma_{2_\mathrm{i}}^2\right)$, where $f_1$ is the flux of the first image, $f_2$ the flux of the second component of the pair, $\sigma_1$ and $\sigma_2$ the corresponding errors. The error $\sigma_\mathrm{R}$ of the estimated R ratio is given by the curvature of the aforementioned sum function at its maximum. The smallest amplification factor A$_\mathrm{min}$ is 1/(R+$\sigma_\mathrm{R}$) and the largest amplification factor A$_\mathrm{max}$ is 1/(R-$\sigma_\mathrm{R}$). Results are reported in Table~\ref{table:lensing}, where the given amplification factor value  is (A$_\mathrm{min}$+A$_\mathrm{max}$)/2 and the error (A$_\mathrm{max}$-A$_\mathrm{max}$)/2.

\begin{table*}
\centering \caption[]{Amplification factors estimated from NACO photometry with a maximum likelihood method (except for pair \#8). Notes: (1) The faintest component is not detected at L'. The maximum likelihood method is used taking 13.5$<$m$_\mathrm{L'}$$<$20 for this star. (2) The first component is only seen in the L'-band and out of the H- and Ks-band fields of view. We use A=$\left(1-10^\mathrm{{\Delta m/2.5}}\right)^{-1}$ where $\Delta$m is firstly the difference between m$_1$-$\sigma$ and m$_2$+$\sigma$, and secondly between m$_1$+$\sigma$ and m$_2$-$\sigma$, with m$_1$ and m$_2$ the L'-band magnitudes of the images and $\sigma$ the photometric error. It leads to say A$_\mathrm{min}$ and A$_\mathrm{max}$. As previously, the given amplification factor  is (A$_\mathrm{min}$+A$_\mathrm{max}$)/2 and the error (A$_\mathrm{max}$-A$_\mathrm{max}$)/2. (3) The identification of the second component is uncertain since the K-band magnitude given in Table~1 of \citet{alexander01} is not the same as in \citet{genzel00}.}
\label{table:lensing}
\resizebox{\hsize}{!}{\begin{tabular}{l c c c c c c c c c c}
\hline
\hline
Pair number & 1 & 2 & 3 & 4 & 5 & 6 & 7 & 8 & 9 & 10\\
\hline
A from NACO & 1.047$\pm$0.005 & 1.96$\pm$0.16 & 1.051$\pm$0.005 & 6.61$\pm$2.65 & 1.045$\pm$0.005 & 3.28$\pm$0.62 & 1.67$\pm$0.10 & 1.23$\pm$0.15 & 1.65$\pm$0.09 & 1.066$\pm$0.008 \\
A from \citet{alexander01} & 1.05 & 3.82 & 1.03 & 2.29 & 1.04 & 2.15 & 2.21 & 1.03 & 1.03 & 1.05 \\
Notes & 1 & & 1 & & 1 & & & 2 & 3 & 1 \\
\hline
\end{tabular}}
\end{table*}

No errors are associated to the amplification factor values in \citet{alexander01}. If one check whether these values are compatible with ours within our given error, only pairs \#1 and \#5 would be possible cases for lensing. Though, pair \#1 is invalidated by spectroscopy \citep{alexander01}. Then, except if one relaxes the uncertainty range, which would sharply decrease the degree of confidence of the result, this photometric study allows one to eliminate all but two pairs of the list.

\section{The IR counterpart to the Northern Arm}

Fig.~\ref{fig:imagelp} shows an extended arc-like structure beginning from the west of \object{IRS~1W}, passing just at the east of \object{IRS~16NE}, through the \object{IRS~16SW} cluster to end just north to \object{IRS~33E}. Known as the Northern Arm (NA), it has been observed at radio wavelength (e.g. \citealp{yusef93}), in near-IR emission lines tracing recombination in the ionized gas: \ion{Ne}{ii} \citep{serabyn88}, Pa $\alpha$ \citep{stolovy99}, Br $\gamma$ \citep{morris00}, \ion{He}{i} \citep{paumard01}.  Because of the large gradient of velocity along the arm, it has been interpreted as a stream of  matter in orbit around the central mass. 

The high angular resolution of our image reveals: a) the extreme narrowness of the brightest part of the arc which is in fact unresolved; b) the decomposition at this scale of  the arc in several arclets, either intersecting or nested, as shown on Fig.~\ref{fig:imagelp}. This picture would not favor the idea of a stream of gas orbiting around \object{Sgr~A*}, since it would not explain a very filamentary concentration nor a departure from a regular elliptical orbit. 

What is the nature of the emission we detect at L' ?  We measure,  for instance, a brightness of 180~mJy~arcsec$^{-2}$ on a typical area of 0.4\arcsec$\times$0.4\arcsec\ on the northern part of the arc. The high frequency counterpart of the free-free emission seen at 15~GHz would correspond to a flux density $\approx$20~mJy~arcsec$^{-2}$, based on the radio brightness of the same typical area  measured on the 15~GHz arc. This is 10 times fainter than our measurement.  On the other hand, if the main contribution to the L'-band emission was the Br $\alpha$ line, the expected power, also scaled from the radio emission, would be 2.4$\times$10$^{-12}$~erg~s$^{-1}$~cm$^{-2}$~arcsec$^{-2}$, which is also 10 times fainter than our value (23$\times$10$^{-12}$~erg~s$^{-1}$~cm$^{-2}$~arcsec$^{-2}$). The additional flux could come from heated dust associated to the gas. The appearance of the  emission, showing up as very thin and possibly broken arclets,  suggests that the arc could be the trace of one or several thin shells of compressed gas due to one or several shocks propagating in the medium:  warm dust within the shocked layer could then become the dominant source of emission in the L'-band. In this scenario, the fact that the velocity is quite different in the northern and southern parts of the arm \citep{morris00} could then be explained if indeed two different shocks are at work.  We are aware that high resolution 10~$\mu$m maps have been obtained at Keck. A comparison of individual features will be done when data is available.

One should note the good matching of the thinest structures between the radio  (Fig.~1 in \citealp{melia01}) and the thermal IR: for instance bright extended filaments south to \object{IRS~2} are clearly observable in Fig~\ref{fig:imagelp} and at centimeter wavelengths.

\section{A new "bow-shock star"}

Several bow-shock stars, interacting with the NA, have been discovered by \citet{tanner02, tanner03} on photometric and radiative transfer modeling basis. Observed with AO at Gemini observatory \citep{rigaut03}, \object{IRS~8} is a bow shock star detected from its morphology, which seems to result from the interaction between a fast moving star and the NA.

Less  than 2\arcsec\ from the NA, 3.44\arcsec\ north and 2.90\arcsec\ east to \object{Sgr A*}, a star shows a bow-shock-like structure on its north-eastern part, in the direction of the NA. The bow-shock figure is composed of 3 "hot spots", each one with a  L'-band surface brightness around 75 mJy~arcsec$^{-2}$, which may indicate a structuration of the interaction. The bow-shock is at a projected distance of 1600~AU from the star. From its absolute L'-band magnitude (M$_\mathrm{L'}$=-5.3) and assuming it is a hot O star, we deduce a star luminosity of about 5$\times$10$^5$~L$_\odot$. The bow shock structure is very red since it appears in the L'-band but neither in the H- nor in Ks-bands: the emission may be warm dust heated by the star. On Fig.~\ref{fig:imagelp}, the thickness of the shell surrounding the star is about 4 pixels, corresponding to 900 AU. Adopting a NA gas density of 10$^5$~cm$^{-3}$ \citep{tanner02}  and a shell depth on the line of sight 2.5 times the thickness, it leads to a column density of N$_\mathrm{H}$=3.4$\times$10$^{21}$~cm$^{-2}$, or  $\tau_\mathrm{V}$$\approx$1.1 and  $\tau_\mathrm{L'}$$\approx$0.06.  In this optically thin case, the observed L'-band brightness ($\approx\tau_\mathrm{L'} \mathrm{B(T_{dust})}$) would  require a dust temperature of 248~K.  From a simple radiative equilibrium model, this is indeed  the temperature reached by silicate dust grains at 1600~AU from a 0.9$\times$10$^5$ L$_\odot$ star at T$_\mathrm{eff}$=40000~K, a quite  reasonable set of parameters.

\section{Summary and conclusions}

We have presented the NACO SV observations of the GC, the IR images with the highest spatial resolution to date. The L'-band image has the highest sensitivity obtained in this wavelength domain. \object{S2}, the closest star to \object{Sgr~A*}, shows a large L'-band excess that may be the first signature of the black hole in the thermal IR. 2003 data already seem to confirm this detection \citep{genzel03b,ghez03b,clenet03} and 2004 observations will probably provide the first spectrum of the IR couterpart of \object{Sgr A*}, \object{S2} being far enough from \object{Sgr~A*} to unambiguously discriminate between each emission. A photometric analysis has allowed us to eliminate all but two of the ten most likely image pair candidates for gravitational lensing given in \citet{alexander01}. This analysis of lensing haven't accounted for any spectroscopic identification of the corresponding stars. Furthermore, the L'-band emission of the NA is seen as narrow intersecting arclets and might trace dust heated by shock compression, which would invalidate precedent interpretations of the NA structure. Finally, a "bow-shock star", interacting with the NA, has been discovered on the L'-band image.  Simulations lead to a dust temperature in the shell of about 250~K. All points discussed above (e.g. nature of \object{S2}, gravitational lensing) will highly benefit from the next spectroscopic high spatial resolution observations scheduled with NACO.

\bibliographystyle{aa}

\end{document}